\documentclass[12pt,floatfix, preprintnumbers]{revtex4}
\usepackage{graphicx}
\usepackage{dcolumn}
\usepackage{bm}
\usepackage{epsfig}
\usepackage{amsmath,amsfonts,amssymb}
\usepackage[outdir=./]{epstopdf}
\graphicspath{ {Images/} }
\usepackage[toc,title]{appendix}
\usepackage{mathrsfs}
\usepackage{float} %

\begin{document}





\title{{\LARGE Single Photon Subradiance:} \\ %
{\normalsize quantum control of spontaneous emission and ultrafast readout} } %
\author{Marlan O. Scully} %
\affiliation{Texas A\&M University, College Station, TX 77843} %
\affiliation{Princeton University, Princeton, NJ 08544} %
\affiliation{Baylor University, Waco, TX 76798} %
\date{\today} %
\begin{abstract}
Recent work has shown that collective single photon emission from an ensemble of resonate two-level atoms, i.e.~single photon superradiance, is a rich field of study. %
The present paper addresses the flip side of superradiance, i.e. subradiance. %
Single photon subradiant states are potentially stable against collective spontaneous emission and can have ultrafast readout.
In particular it is shown how many atom collective effects provide a new way to control spontaneous emission %
by preparing and switching between subradiant and superradiant states. %
\end{abstract}
\maketitle

\newpage

\begin{figure}[th]
\centering
\includegraphics[width=0.75\textwidth]{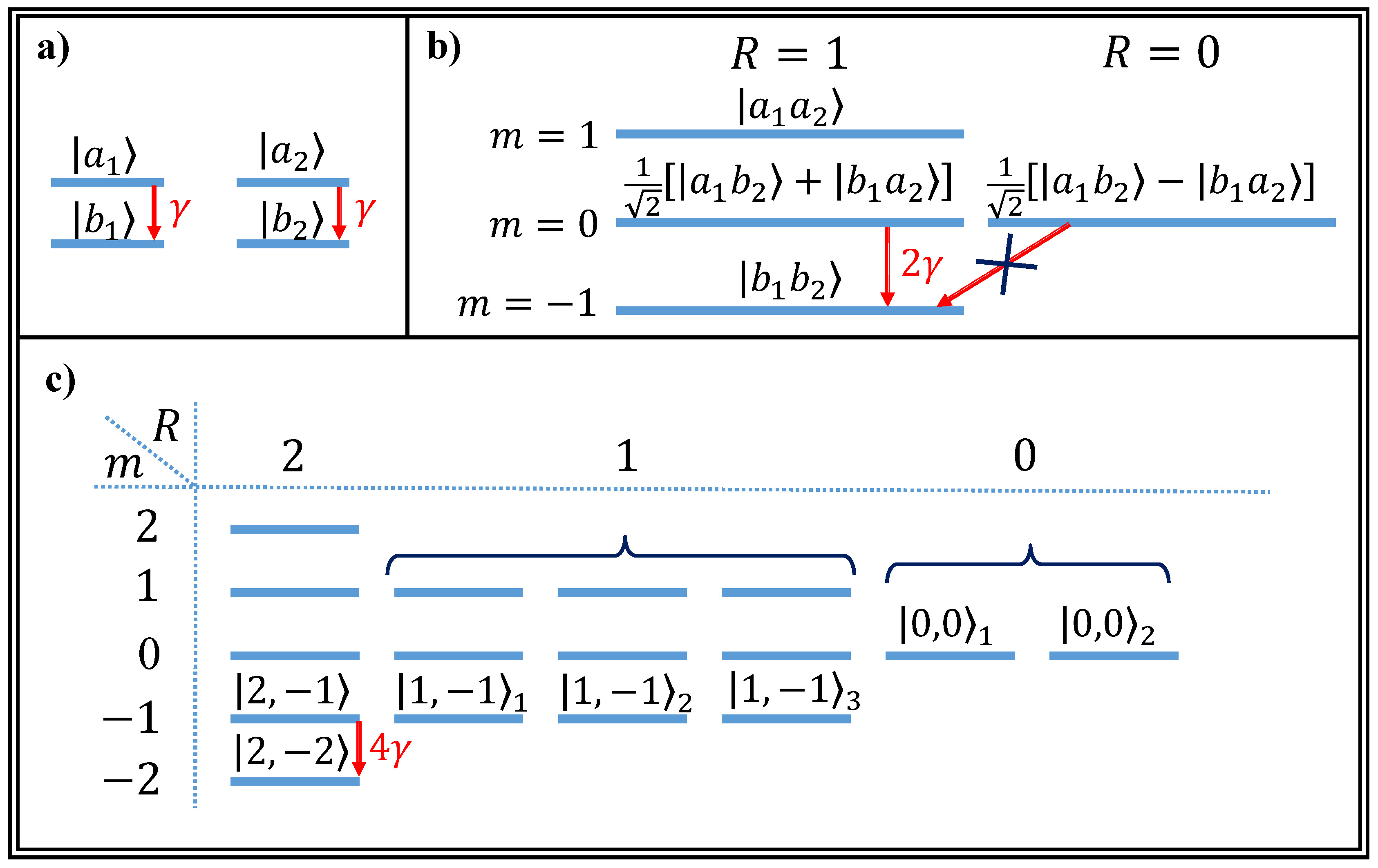} %
\caption{a) The upper (lower) states for atoms $j = 1, 2$ are $\vert a_j \rangle$ ($\vert b_j \rangle$). %
b) The two atoms can be grouped into triplet $R = 1$ and singlet $R = 0$ state, %
where $R = N/2$ is the cooperation number, %
and $m$ is $(N_a - N_b) / 2$ where $N_a$ and $N_b$ are the number of atoms in $\vert a \rangle$ and $\vert b \rangle$. %
c) The group characterization of the 16 atom states is depicted in an $\vert R, m \rangle_p$ notation %
where the $p$ index labels the column within a given $R$. %
The states $R = 1$ and $R = 0$ are $3$ and $2$ fold degenerate. %
The superradiant state $\vert 2, -1 \rangle$ decays at a rate four times that of a single atom. %
The states $\vert 1, -1 \rangle_s$ ($s = 1,2,3$) contain one photon energy and do not decay. %
It may be noted for $R = 1$ the $s$ index denotes the number of singlet states, see e.g. Table~\ref{Table:SubradiantStates}. %
The states $\vert 0, 0 \rangle_1$ and $\vert 0, 0 \rangle_2$ are two photon subradiant states %
involving $1$ and $2$ \underline{pairs} of singlet states. %
} %
\label{Fig:LevelStructureStates} 
\end{figure}

Group theory is one of the most beautiful subjects in physical mathematics. %
No better example than Dicke superradiance \cite{Dicke1954,Burnham1969PR,Friedberg1973,Gross1982,Prasad2000}. %
Dicke taught us that radiating two level atoms can be insightfully grouped into angular momentum multiplets. %
To motivate this connection we note that each two level atom is a spinor. %
Then for two atoms (or two neutrons in a magnetic field, etc.) %
it takes four states to cover the spin space. %
The four spin states can be grouped into the spin triplet and singlet states depicted in the upper right corner of Fig.~\ref{Fig:LevelStructureStates}. %
As an example of the utility of the method the decay states of the system can now be read off using angular momentum matrix elements. %
Indeed the term ``superradiance'' derives from the fact that the decay rate from state $\vert r, m \rangle$ to $\vert r, m-1 \rangle$ %
goes like the square of the angular momentum lowering operator connecting these states which is given by $(r+m) (r-m+1)$. %
Hence when $m = 0$ (equal population in the ground and excited states) the radiation emission rate goes as $N^2$. %
This superradiant rate can be understood semiclassically by noting that the electric field emitted by $N$ coherently prepared dipoles is proportioned to $N$, %
and the intensity goes as $N^2$. %
In fact most of the calculations associated with superradiance experiments are carried out using a semiclassical Maxwell-Bloch formalism.

However, recent work has shown that collective single photon spontaneous emission from an ensemble of resonate two-level atoms %
is a rich field of study %
\cite{Scully2006,Mazets2007,Akkermans2008,Wiegner2011,WeiFeng2014,DaWei2015}. %
For example single photon superradiance from an ensemble much larger than the radiation wavelength %
yields enhanced directional spontaneous emission; %
and when the ensemble becomes larger than the radiation wave packet \cite{Scully2009Science} %
interesting collective superradiant effects abound. %

As is stated in the abstract and discussed below following, Eq.~\eqref{Eq:MinusState}, %
the present focus is on control of spontaneous emission and the switching from subradiant to superradiant states. %
Application to single photon devices is apparent, but the study of cooperative subradiance is of interest in and of itself. %

\begin{figure}[ht] %
\centering %
\includegraphics[width=0.75\textwidth]{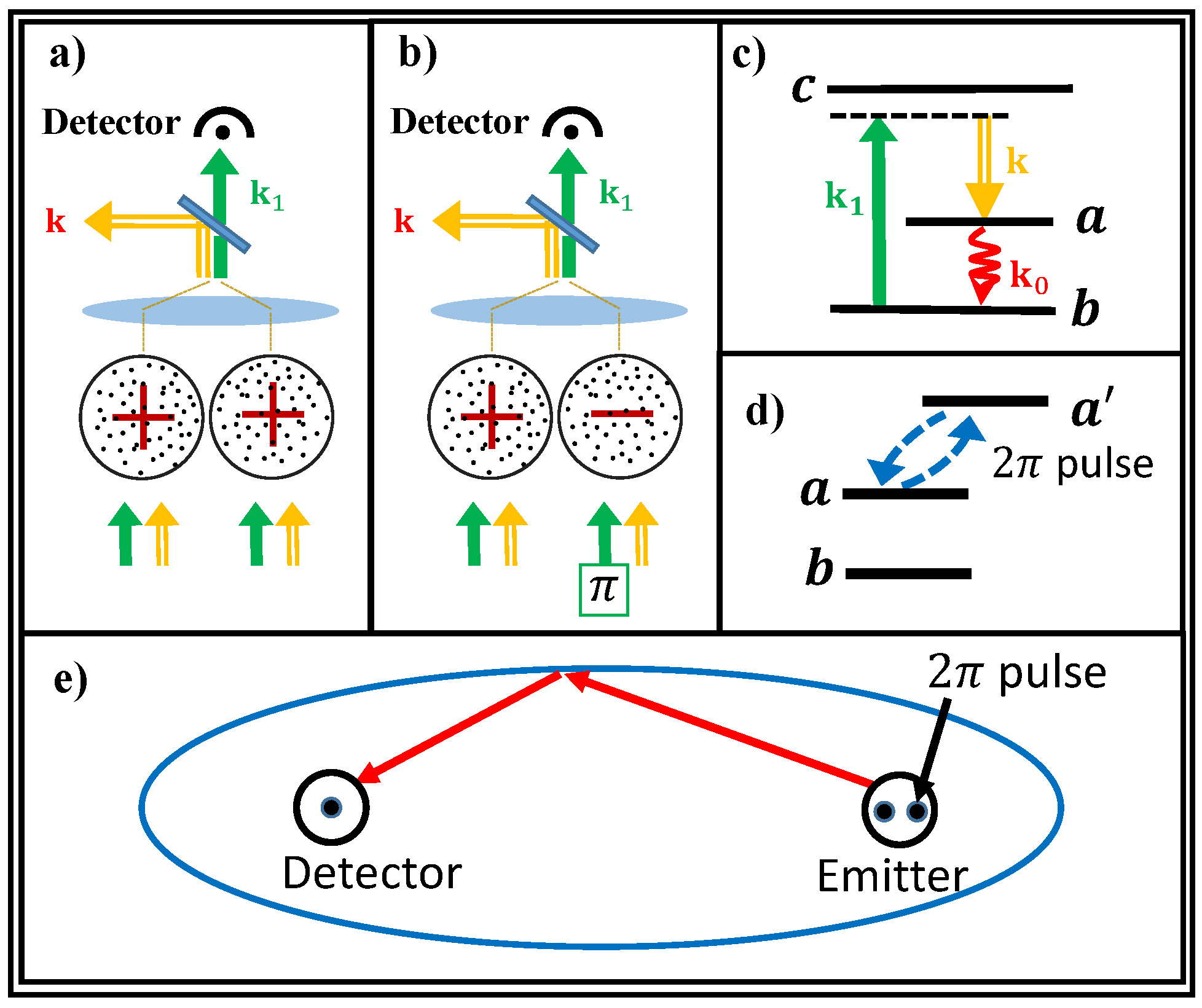}
\caption{a) Excitation of $\vert + \rangle_N$ state. %
A single photon of wave vector $\mathbf{k}_1$ is accompanied by a laser having wave vector $\mathbf{k}$, %
$\mathbf{k} - \mathbf{k}_1 = \mathbf{k}_0$. %
The $\mathbf{k}_0$ photon is resonant with the transition  $\vert a \rangle$ to $\vert b \rangle$. %
The atoms are weakly driven by the excitation process i.e.~the atomic medium is optically thin during the preparation process, %
and most of the $\mathbf{k}_1$ single photon pulses do not register a count in the detector; %
the $\mathbf{k}$ radiation is isolated from the detector. %
The lack of a count heralds the preparation of the $\vert + \rangle_N$ state. %
b) Same as part (a) but single photon $\mathbf{k}$ is divided by a beam splitter %
and shifted by $\pi$ on the RHS so those atoms are prepared out of phase with the LHS atoms.
That is the $\vert a_{j'} \rangle$ atoms on the RHS are multiplied by $-1$; %
the net result in (a) and (b) is that a no-count event signals the fact that the $\vert \pm \rangle_N$ state has been prepared. %
c) The atoms are weakly driven by a Raman-type process in which two photons $\mathbf{k}$ and $\boldsymbol{\kappa}$ excite the atom to a virtual state 
and the $\mathbf{k}_1$ photon %
takes the atom to the $\vert a \rangle$ state. %
d) Cycling the RHS atoms $a \rightarrow a' \rightarrow a$ results in another factor of $-1$, which when applied to the RHS atoms %
takes the single photon subradiant state $\vert - \rangle$ to $\vert + \rangle$. %
e) Sketch of double microdot emitter which is small compared to the superradiant wavelength ($\lambda_0 > 1 \mu$). %
As in Fig.~\ref{Fig:SubradiantState}b the double dots are initially prepared in $\vert - \rangle_N$; %
and then after some storage time ($T > \text{millisec}$) switched to the $\vert + \rangle_N$ state by the $2\pi$ pulse which drives the RHS dot %
as depicted in Fig.~\ref{Fig:SubradiantState}d. %
The ellipsoidal cavity directs all $\lambda_0$ photons emitted to the detector. %
The cavity walls are transparent to the preparation radiation $\mathbf{k}_1$ and $\mathbf{k}$ of Fig.~\ref{Fig:SubradiantState}c as well as the $2\pi$ switching pulses. %
} %
\label{Fig:SubradiantState} %
\end{figure}

The physics of the subradiant states was explained by Dicke \cite{Dicke1954}. %
These states are necessary to span the $N$ atom space. %
For example, the subradiant states are used in calculating the many atom Lamb shift via the timed Dicke states \cite{Scully2009Science},
but they have not stimulated anything like the amount of work that the superradiant state has. %
to be sure, interesting work on Dicke subradiance has been reported. %
For example, Pavolini and coworkers \cite{Pavolini1984PRL} observed a reduced emission rate due to ``trapping'' of a fraction of the atoms in a mixture of many subradiant states following pulsed excitation. %
Some experiments of interest have focused on preparation of the subradiant state of a two atom system \cite{DeVoe1996PRL}. %
Other interesting work \cite{Kaiser2012PRL} on an $N$ atom inhomogeneously broadened ensemble driven by a weak pulse %
has demonstrated about $0.1 \%$ in a mixture of subradiant states. %
Subradiance in molecular \cite{McGuyer2014NaturePhysics} and quantum dot \cite{Temnov2005PRL} systems is also of interest. %
We here present and analyze a simple method whereby a substantial fraction, in some cases even $100 \%$, of the atoms can be placed in the single $N$ atom subradiant state %
given by Eq.~\eqref{Eq:MinusState}. %
Furthermore, it is possible to switch between the subradiant state of Eq.~\eqref{Eq:MinusState} %
to the companion superradiant state of Eq.~\eqref{Eq:RadiantStates_Dicke} by $2\pi$ cycling of half the atoms %
as depicted in Fig.~\ref{Fig:SubradiantState}. %

For our purposes it takes at least four atoms to introduce the story. %
The $2^4$ states of the four spin system of Fig.~\ref{Fig:LevelStructureStates} and Table 1 are spanned by the angular momentum multiplets %
having total ``angular momentum'' $R = 2$, $1$ and $0$. %

\begin{table}[ht] %
\begin{tabular}{c|c} 
\hline \hline %
Subradiant & Singlet State \\ %
State & Representation \\ %
\hline %
$\vert 1, -1 \rangle_1$ & $\vert s_{12} \rangle \vert b_3 b_4 \rangle$ \\ %
\hline
$\vert 1, -1 \rangle_2$ & $\big[ \vert s_{13} \rangle \vert b_2 b_4 \rangle + \vert s_{23} \rangle \vert b_1 b_4 \rangle \big]/ \sqrt{3}$ \\ %
\hline
$\vert 1, -1 \rangle_3$ & \, $\big[ \vert s_{14} \rangle \vert b_2 b_3 \rangle + \vert s_{24} \rangle \vert b_1 b_3 \rangle %
+ \vert s_{34} \rangle \vert b_1 b_2 \rangle \big] / \sqrt{6}$ \quad \\ %
\hline
$\vert 0, 0 \rangle_1$ & \, $\vert s_{12} \rangle \vert s_{34} \rangle$ \quad \\ %
\hline
$\vert 0, 0 \rangle_2$ & \, $\big[ \vert s_{13} \rangle \vert s_{24} \rangle + \vert s_{23} \rangle \vert s_{14} \rangle \big] / \sqrt{3}$ \quad \\ %
\hline \hline %
\end{tabular} %
\caption{Subradiant states for a 4-atom system in terms of 2 atom singlet states, %
where $\vert s_{ij} \rangle = [ \vert a_i b_j\rangle - \vert b_i a_j \rangle ] / \sqrt{2}$. %
The notation is explained in Fig.~1. %
See also Supplement B. %
} %
\label{Table:SubradiantStates} %
\end{table} %

As discussed in the preceeding, %
we seek single photon subradiant states which are long lived and can be switched to the single photon superradiant state of Eq.~\eqref{Eq:MinusState}
given by %
\begin{subequations} %
\label{Eq:RadiantStates_Dicke} %
\begin{align} %
\label{Eq:Superradiant_Dicke} %
\vert + \rangle_N = \frac{1}{\sqrt{N}} \sum_{j=1}^N \vert j \rangle, %
\end{align} %
where $\vert j \rangle = \vert b_1 \cdots a_j \cdots b_N \rangle$ and $b_j (a_j)$ is the ground (excited) state of the $j$th atom. %
The companion subradiant state of Fig.~\ref{Fig:SubradiantState}b is given by %
\begin{align} %
\label{Eq:Subradiant_Dicke} %
\vert - \rangle_N = \frac{1}{\sqrt{N}} \left[ \sum_{j=1}^{N/2} \vert j \rangle \:\: - \sum_{j'=N/2+1}^N \vert j' \rangle \right]. %
\end{align} %
\end{subequations} %

In the following we first investigate single photon superradiant in small and extended samples using single photon preparation. %
And, as is shown in Supplement A, the $\vert + \rangle_N$ state decays at the superradiant state rate $N\gamma$, %
where $\gamma$ is the single atom decay rate and the $\vert - \rangle_N$ state is long lived. %
In both cases post selection allows for a thin optical medium on preparation and a thick medium on decay. %
Subradiant state preparation of a four atom system %
to state $\vert 0, 0 \rangle$ and to $N$ atom super- and sub-radiant states Eq.~(3a,b) %
without post selection is also presented. %
We conclude with a brief summary of results and open questions. %

It is useful to write the $\vert - \rangle_N$ state in terms of the $\vert s_{jj'} \rangle$ singlet states of Table \ref{Table:SubradiantStates}, where the $j$ index runs from $1$ to $N/2$ %
and $j'$ runs from $N/2+1$ to $N$; that is %
\begin{align} %
\label{Eq:MinusState} %
\vert - \rangle_N = \frac{1}{\sqrt{N/2}} \sum_{j,j'} \vert s_{jj'} \rangle \vert \{ b \}_{jj'} \rangle, %
\end{align} %
where $\vert \{ b \}_{jj'} \rangle$ is the $N$ atom ground state with the $j$ and $j'$ atoms removed. 
Having stored a photon in state $\vert - \rangle_N$ we can extract, %
i.e.~readout this information by switching the minus sign in Eq.~\eqref{Eq:Subradiant_Dicke} %
to a plus. %
This we do by cycling the RHS $j'$ atoms with a $2\pi$ pulse as in Fig.~\ref{Fig:SubradiantState}d. %
This results in $\vert a_{j'} \rangle$ going to $- \vert a_{j'} \rangle$ %
and we change the subradiant state Eq.~\eqref{Eq:Subradiant_Dicke} to the symmetrical state $\vert + \rangle$ given by Eq.~\eqref{Eq:Superradiant_Dicke}, %
which decays at the rate $\Gamma_N = N\gamma$. %
Thus the single photon subradiant state $\vert - \rangle$ is stable against collective spontaneous emission %
with storage time long compared with $\Gamma_N^{-1}$. %
And can be addressed in ultrashort switching times of order $(N\gamma)^{-1}$ ($\lesssim \text{picosec}$), see Fig.~\ref{Fig:SubradiantState}e. %

By way of comparison a photon stored in a high $Q$ cavity has a lifetime of several $\mu\text{sec}$, %
and the cavity switching times as determined electronically are usually in the nanosecond range. %
Atomic dark states can also store a photon for long times %
but the switch out times are of order $\gamma^{-1}$. %

These results have an extension to the large sample (timed Dicke) case. %
Then, as is shown in \cite{Scully2006}, %
the superradiant state prepared by a photon of vector $\mathbf{k}_0$ is given by %
\begin{subequations} %
\begin{align} %
\label{Eq:SuperradiantState_SinglePhoton} %
\vert + \rangle_{\mathbf{k}_0} = \frac{1}{\sqrt{N}} \sum_{j=1}^{N} e^{i {\mathbf{k}_0} \cdot \mathbf{r}_j} \vert b_1 b_2 \cdots a_j \cdots b_N \rangle, %
\end{align} %
and as discussed in Supplement B the corresponding single photon subradiant state is given by %
\begin{align} %
\label{Eq:Subradiant_State} %
\vert - \rangle_{\mathbf{k}_0} = & \, \frac{1}{\sqrt{N}_2} {\sum_{j,j'}}^{2} \frac{1}{\sqrt{2}} \left( \vert a_j b_{j'} \rangle e^{i{\mathbf{k}_0} \cdot \mathbf{r}_j} %
- \vert b_j a_{j'} \rangle e^{i{\mathbf{k}_0} \cdot \mathbf{r}_{j'}} \right) %
\, \vert \{ b \}_{jj'} \rangle. %
\end{align} %
\end{subequations} %
Then, the $\vert + \rangle_{\mathbf{k}_0}$ state decays approximately at the rate %
\begin{subequations} %
\begin{align} %
\Gamma_+ \cong \frac{3}{16\pi} \gamma \frac{\lambda^2}{A} (N-1) + \frac{\gamma}{2}, %
\end{align} %
and the corresponding $\vert - \rangle_{\mathbf{k}_0}$ state decay rate goes as %
\begin{align} %
\Gamma_- \cong \frac{3}{16\pi} \gamma \frac{\lambda^2}{A} \left( \frac{N}{2} - \frac{N}{2} \right) + \frac{\gamma}{2} - \frac{3}{16\pi} \gamma \frac{\lambda^2}{A}. %
\end{align} %
\end{subequations} %
It is important to note that $\Gamma_-$ is lower bounded by the natural single atom rate $\gamma$. %
But the collective decay rates $\Gamma_+$ and $\Gamma_-$ are very different %
and one can envision cases where a storage time of order $\gamma^{-1}$ with directional photons emitted %
in times of order $\Gamma_+^{-1}$ would be of interest. %
However in the interest of simplicity, the focus of the present paper is the preparation of the $\vert - \rangle_N$ state %
and subsequent switching between the single photon subradiant states, neighboring single photon superradiant and two photon subradiant states. %

As an aside we note that higher energy states (e.g. $\vert 1, 0 \rangle_1$ of Fig.~\ref{Fig:LevelStructureStates}c) can be realized %
by driving the system with the symmetric raising operator $\hat{R}^+ = \sum_j \hat{\sigma}_j^+$. %
In particular the $\vert - \rangle_N$ state of Eq.~\eqref{Eq:MinusState} is promoted to the two photon state %
\begin{align} %
\vert - \rangle^{(2)}_N = \frac{1}{\sqrt{N/2}} \sum_{j,j'} \vert s_{jj'} \rangle \vert +_{jj'} \rangle %
\end{align} %
where $\vert +_{jj'} \rangle$ is the symmetric state Eq.~\eqref{Eq:RadiantStates_Dicke} with $j$ and $j'$ atoms missing. %
This example makes clear the utility of the writing the subradiant states %
in the $\vert s_{jj'} \rangle \vert \{ b \}_{jj'} \rangle$ form of Eq.~\eqref{Eq:MinusState}, %
namely the raising $\hat{R}^+$ operator only acts on the $\vert \{ b \}_{jj'} \rangle$ states %
since $\hat{R}^+ \vert s_{jj'} \rangle = 0$. %

\begin{figure}[ht] %
\centering %
\includegraphics[width=0.7\textwidth]{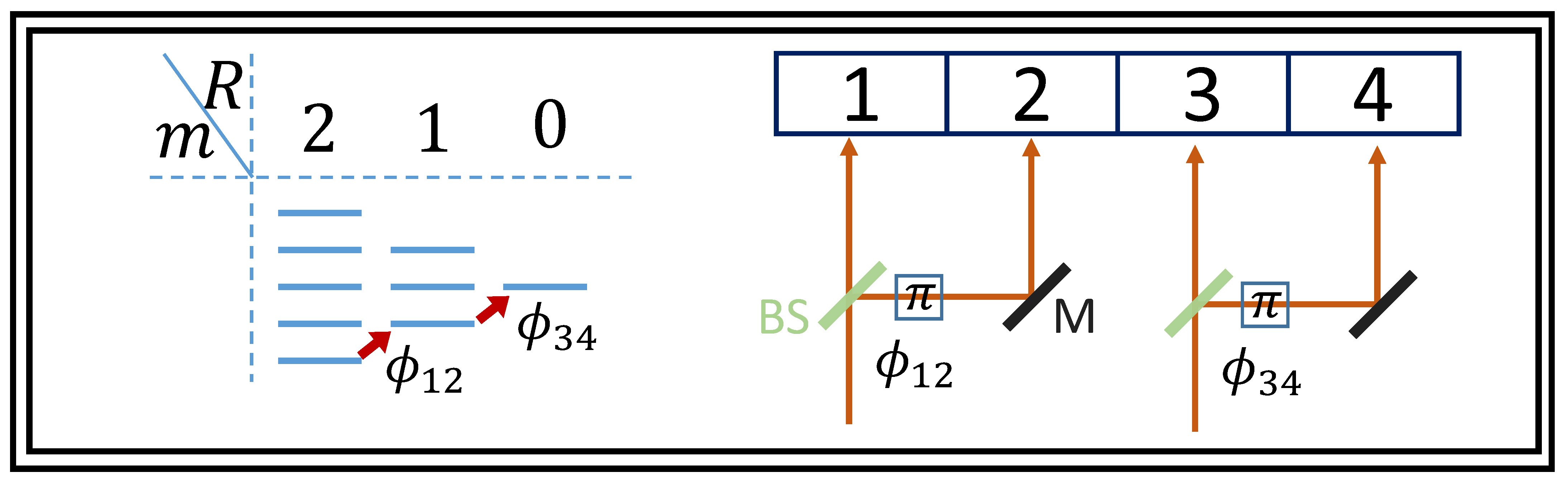} %
\caption{ %
The single photon subradiant state $\vert 1, -1 \rangle_1$ is prepared by coupling atoms $1$ and $2$ in $\vert 2, -2 \rangle = \vert b_1 b_2 b_3 b_3 \rangle$ %
with the single photon state $\vert \Phi \rangle_{12}$ of Eq.~\eqref{Eq:InitialPhotonState} prepared via a $\pi$ phase shift as indicated. %
The double photon subradiant $\vert 0, 0 \rangle_1 = \vert s_{12} s_{34} \rangle$ is prepared %
by coupling atoms $3$ and $4$ in $\vert 1, -1 \rangle_1 = \vert s_{12} \rangle \vert b_3 b_4 \rangle$ with $\vert \Phi \rangle_{34}$. %
$BS$ and $M$ denotes beam splitters and mirrors and \fbox{$\pi$} indicates a $\pi$ phase shifter. %
} %
\label{Fig:MultiPhotonSubradiant} %
\end{figure}

Next we turn to the four atom case of Fig.~\ref{Fig:LevelStructureStates} %
and consider preparation of single photon $\vert 1, -1 \rangle_1$ and two photon $\vert 0, 0 \rangle_1$ subradiant states.
Consider the four atom system in Fig.~\ref{Fig:MultiPhotonSubradiant}. %
There we see the four atom states $\vert 1, -1 \rangle_1$ and $\vert 0, 0 \rangle_1$ prepared by sidewise illumination. %
In particular, the single photon subradiant state %
$\vert 1, -1 \rangle_1$ %
is prepared by passing a single photon $\vert \gamma \rangle$ through a beam splitter, %
and phase shifting the radiation directed to atom $2$ by $\pi$ as in Fig.~\ref{Fig:MultiPhotonSubradiant}. %
The light in the two legs of the optical path interacts with atoms $1$ and $2$ %
and post-selecting the photon vacuum %
then prepares the singlet state $\vert s_{12} \rangle \vert b_3 b_4 \rangle$. %
That is, beginning with the photon state %
\begin{align} %
\label{Eq:InitialPhotonState} %
\vert \Phi \rangle_{12} = \frac{1}{\sqrt{2}} \left( \vert 1_1, 0_2 \rangle - \vert 0_1, 1_2 \rangle \right), %
\end{align} %
where atoms $1$ and $2$ are driven by photons $\vert 1_1 \rangle$ and $\vert 1_2 \rangle$ via the resonant interaction %
\begin{align} %
\nonumber %
V = \frac{1}{2} \hbar g \sum_{i=1,2} ( \hat{\sigma}_i^+ \hat{a}_i + \hat{a}_i^\dagger \hat{\sigma}_i ) %
\end{align} %
where $\hbar g$ is the product of the atomic matrix element $\wp$ and the electric field per photon $\mathcal{E} = \sqrt{\hbar \nu / \epsilon_0 V}$ %
where $\nu$ is the photon frequency. %
Then for both systems the atom field state evolves according to $U(t) = \exp (- i Vt/\hbar)$; %
and one finds %
\begin{subequations} %
\label{Eq:EvolutionOperatorW} %
\begin{align} %
\label{Eq:EvolutionOperator} %
U(t) \vert b, 1 \rangle = \cos \left( \frac{1}{2} g t \right) \vert b, 1 \rangle %
- i \sin \left( \frac{1}{2} g t \right) \vert a, 0 \rangle. %
\end{align} %
Hence if $g\tau = \pi$, i.e.~we have a single photon $\pi$ pulse, then $\vert b, 1 \rangle \rightarrow \vert a, 0 \rangle$ for both atoms $1$ and $2$; %
thus from Eq.'s \eqref{Eq:InitialPhotonState} and \eqref{Eq:EvolutionOperator} we find %
\begin{align} %
U_1(\tau) U_2(\tau) \vert b_1 b_2 \rangle \frac{1}{\sqrt{2}} \left[ \vert 1_1 0_2 \rangle - \vert 0_1 1_2 \rangle \right] %
= \frac{1}{\sqrt{2}} [ \vert a_1 b_2 \rangle - \vert b_1 a_2 \rangle ] \vert 0_1 0_2 \rangle. %
\end{align} %
\end{subequations} %
Likewise atoms $3$ and $4$ can be prepared in the singlet state $\vert s_{34} \rangle$ and we arrive at the double singlet state %
$\vert 0, 0 \rangle_1 = \vert s_{12} \rangle \vert s_{34} \rangle$ without doing any post selection. %

\begin{figure}[ht] %
\centering
\includegraphics[width=0.8\textwidth]{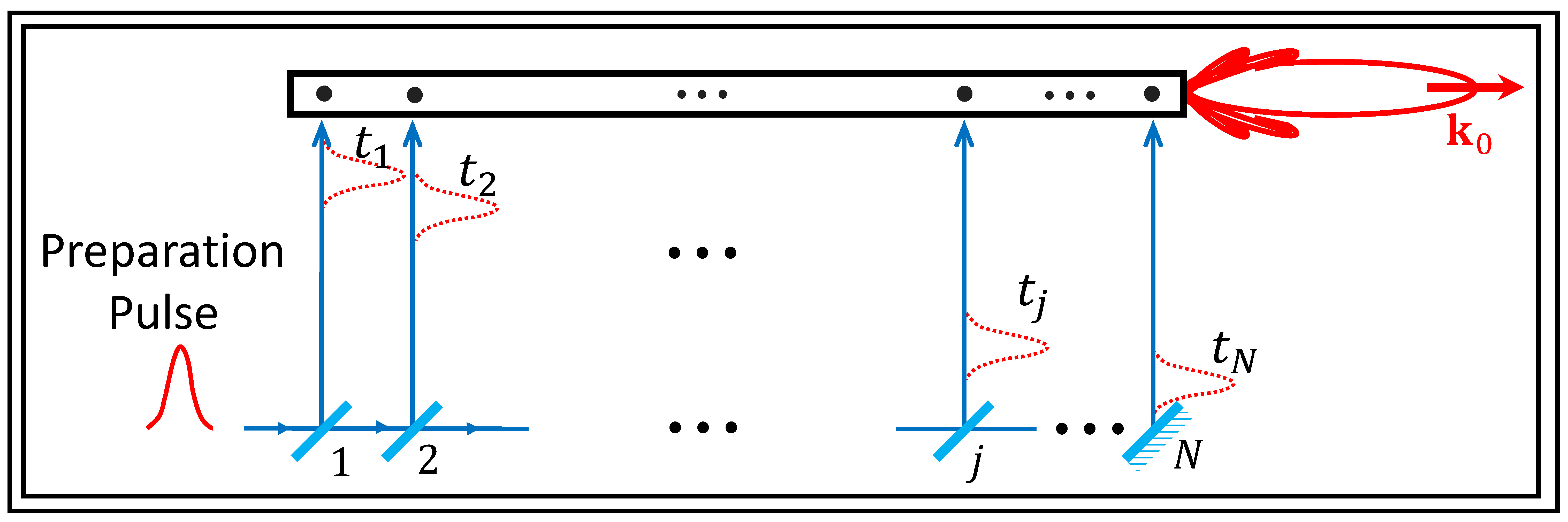} %
\caption{ %
Sidewise excitation of extended medium by single photon which passes through a series of beam splitters and is then directed onto an array of atoms %
orthogonal to the emission direction. %
The single photon preparation pulse is a $\pi$ pulse which drives the atoms from $\vert b \rangle$ to $\vert a \rangle$. %
} %
\label{Fig:SidewiseExcitation} %
\end{figure}

It is also interesting to note that we can use the sidewise excitation scheme of Fig.~3 to prepare an $N$ atom timed Dicke state %
without post selection. %
This we do by passing a single photon $\pi$ pulse through a series of beam splitters (BS's) arranged as in Fig.~\ref{Fig:SidewiseExcitation}. %
There we see a series of atoms in known fixed positions correlated with beam splitters with varying reflectance. %
That is, the 1st BS has a reflectance $r_1 = 1/\sqrt{N}$ and the $N^\text{th}$ BS is a mirror with $r_N = 1$; %
each BS is selected so that a field of strength $E_\text{in} / \sqrt{N}$ is focused on the atoms depicted in Fig.~\ref{Fig:SidewiseExcitation}. %
For example, for three atoms $r_1 = 1/\sqrt{3}$, $r_2 = \sqrt{2/3}$, and $r_3 = 1$. %
Thus a single photon state $\vert 1_{\mathbf{k}_0} \rangle$ injected from the left will be split into $N$ modes %
so that the $N$-mode photon/$N$ atom system given by %
\begin{subequations} %
\begin{align} %
\vert \Psi(0) \rangle = \frac{1}{\sqrt{N}} \sum_j e^{ik_0 z_j} \vert 0_1 0_2 \cdots 1_j \cdots 0_N \rangle \vert b_1 b_2 \cdots b_j \cdots b_N \rangle %
\end{align} %
evolves (for single photon $\pi$ pulse excitation) according to Eq.~\eqref{Eq:EvolutionOperatorW} into %
\begin{align} %
\vert \Psi(t) \rangle = \frac{1}{\sqrt{N}} \sum_j e^{ik_0 z_j} \vert b_1 \cdots a_j \cdots b_N \rangle \vert \{ 0 \} \rangle %
\end{align} %
\end{subequations} %
which is the $\vert + \rangle_{\mathbf{k}_0}$ state of Eq.~\eqref{Eq:SuperradiantState_SinglePhoton}. %
Likewise a $\pi$ phase shifter placed between the $\tfrac{1}{2} N$ and $\tfrac{1}{2}N+1$ BS will yield the $\vert - \rangle_{\mathbf{k}_0}$ state given by Eq.~\eqref{Eq:Subradiant_State}. %

By way of summary and open questions: %
the control of spontaneous emission is a problem of long-standing interest. %
For example, storing an excited atom in a cavity detuned from atomic resonance will slow atomic decay. %
Then upon switching the cavity into atomic resonance the atom will decay. %
This is possible on e.g.~a microsecond (cavity decay time) scale. %
In the present scheme we can potentially hold off spontaneous emission for a much longer time; %
and then switch from subradiance to superradiance which can produce emission in e.g.~a picosecond time scale. %
Sample preparation will be challenging likely involving micron cryogenic color center or nitrogen vacancy diamond quantum dots. %
Larger configurations involving timed Dicke states or their extensions %
(e.g.~placing the atoms at periodic sites, such that the set $\{ k_0 z_j \}$ is tailored appropriately), %
can also be useful %
and will be reported elsewhere, see also Supplement A. %
Other single photon subradiant states are a natural extension of the present approach. %
See e.g.~the three sub-ensemble state depicted in Fig.~B2. %
\begin{align} %
\widetilde{\vert - \rangle} = \frac{1}{\sqrt{6}} \left[ \sum_{j=1}^{N/3} \frac{e^{i\mathbf{k}_0 \cdot \mathbf{r}_j}}{\sqrt{ N/3}} \vert j \rangle %
-2 \sum_{j'=N/3+1}^{2N/3} \frac{e^{i\mathbf{k}_0 \cdot \mathbf{r}_{j'}}}{\sqrt{ N/3}} \vert j' \rangle %
+ \sum_{j''=2N/3+1}^{N} \frac{e^{i\mathbf{k}_0 \cdot \mathbf{r}_{j''}}}{\sqrt{ N/3}} \vert j'' \rangle \right]. %
\end{align} %
In particular we note that the sidewise excitation uses the single photon preparation pulses efficiently as compared to conditional excitation. %
It is also interesting that a semiclassical treatment fails in the scheme of Fig.~\ref{Fig:SidewiseExcitation}. %
This will be further discussed elsewhere. %

The effects of the many particle cooperative Lamb shift %
\cite{Friedberg2008PhysLettA,Rohlsberger2010,Keaveney2012PRL,Meir2014PRL} %
have been found to be interesting in the timed Dicke single photon superradiant state. %
Likewise the Lamb shift in the many particle subradiant states is an interesting problem as is %
the Agarwal-Fano coupling between collections of single photon super- and sub-radiant states, %
and will be the subject of further studies. %
Multiphoton subradiant states (e.g.~$\vert 0, 0 \rangle_1$ and $\vert 0, 0 \rangle_2$ of Fig.~1) %
suggest interesting open questions, involving preparation of these states using classical fields. %
In general, single photon subradiance, its preparation and manipulation provides many open questions. %

\textbf{Acknowledgements:} I thank B.~Kim, H.~Cai, H.~Dong, I.~Mirza, W.~Schleich, G.~Shchedrin, A.~Sokolov, A.~Svidzinsky, D.~Wang, and L.~Wang for discussions, %
and NSF Grant PHY-1241032 and Robert A.~Welch Foundation Award A-1261 for support. %


\begin{thebibliography}{90}

\bibitem{Dicke1954} %
R.H.~Dicke, Phys. Rev. \textbf{93}, 99 (1954). %

\bibitem{Burnham1969PR} %
D.C.~Burnham and R.Y.~Chiao, Phys.~Rev.~\textbf{188}, 667 (1969). %

\bibitem{Friedberg1973} %
R.~Friedberg, S.R.~Hartmann, and J.T.~Manassah, Phys.~Rep.~\textbf{7}, 101 (1973). %

\bibitem{Gross1982} %
M.~Gross and S.~Haroche, Phys.~Rep.~\textbf{93}, 301 (1982). %

\bibitem{Prasad2000} %
S.~Prasad and R.J.~Glauber, Phys. Rev. A \textbf{61}, 063814 (2000). %

\bibitem{Scully2006} %
M.O.~Scully, E.S.~Fry, C.H.Raymond Ooi, and K.~W\'odkiewicz, Phys. Rev. Lett. \textbf{96}, 010501 (2006). %

\bibitem{Mazets2007} %
I.~Mazets and G.~Kurizki, J.~Phys.~B: At.~Mol.~Opt.~Phys.~\textbf{40}, F105 (2007). %

\bibitem{Akkermans2008} %
E.~Akkermans, A.~Gero, and R.~Kaiser, Phys.~Rev.~Lett.~\textbf{101}, 103602 (2008). %

\bibitem{Wiegner2011}
R.~Wiegner, J.~von Zanthier, and G.S.~Agarwal, Phys.~Rev.~A \textbf{84}, 023805 (2011). %

\bibitem{WeiFeng2014} %
W.~Feng, Y.~Li and S.Y.~Zhu, Phys.~Rev.~A \textbf{89}, 013816 (2014). %

\bibitem{DaWei2015} %
D.-W.~Wang, H.~Cai, L.~Yuan, S.Y.~Zhu, and R.B.~Liu, Optica \textbf{2}, 712 (2015).

\bibitem{Scully2009Science} %
M.O.~Scully and A.~Svdzinsky, Science \textbf{325}, 1510 (2009); M.O.~Scully, Phys.~Rev.~Lett.~102, 143601 (2009). %

\bibitem{Pavolini1984PRL} %
D.~Pavolini, A.~Crubellier, P.~Pillet, L.~Cabaret, and S.~Liberman, Phys.~Rev.~Lett.~\textbf{54}, 1917 (1985). %

\bibitem{DeVoe1996PRL} %
R.G.~DeVoe and R.G.~Brewer, Phys.~Rev.~Lett.~\textbf{76}, 2049 (1996). %

\bibitem{Kaiser2012PRL} %
T.~Bienaim\'e, N.~Piovella, and R.~Kaiser, Phys.~Rev.~Lett.~\textbf{108}, 123602 (2012). %

\bibitem{McGuyer2014NaturePhysics} %
B.H.~McGuyer, M.~McDonald, G.Z.~Iwata, M.G.~Tarallo, W.~Skomorowski, R.~Moszynski and T.~Zelevinsky, Nature Physics \textbf{11}, 32 (2015). %

\bibitem{Temnov2005PRL} %
V.V.~Temnov and U.~Woggon, Phys.~Rev.~Lett.~\textbf{95}, 243602 (2005). %

\bibitem{Friedberg2008PhysLettA} %
R.~Friedberg and J.T.~Manassah, Phys.~Lett.~A \textbf{372}, 2514 (2008).

\bibitem{Rohlsberger2010} %
R.~R\"ohlsberger, K.~Schlage, B.~Sahoo, S.~Couet, R.~R\"uffer, Science \textbf{328}, 1248 (2010). %

\bibitem{Keaveney2012PRL} %
J.~Keaveney, A.~Sargsyan, U.~Krohn, I.G.~Hughes, D.~Sarkisyan, and C.S.~Adams, Phys.~Rev.~Lett.~\textbf{108}, 173601 (2012). %

\bibitem{Meir2014PRL} %
Z.~Meir, O.~Schwartz, E.~Shahmoon, D.~Oron, and R.~Ozeri, Phys.~Rev.~Lett.~\textbf{113}, 193002 (2014). %

\end{thebibliography}
\end{document}






\begin{center} %
\large Supplements to ``Single Photon Subradiance: quantum control of spontaneous emission and ultrafast readout'' %
\end{center} %


\begin{appendices}

\section*{Supplement A. Weisskopf-Wigner treatment of radiative decay from single photon super- and sub-radiant states} %
\renewcommand{\thesection}{\Alph{section}}
\setcounter{section}{1}
\setcounter{equation}{0}
\numberwithin{equation}{section}
\renewcommand{\theequation}{\thesection.\arabic{equation}a}

Consider first the simple case of decay of the $\vert + \rangle$ and $\vert - \rangle$ states in the small sample Dicke limit. %
The state %
\begin{align} %
\vert \Psi(t) \rangle = %
\beta_+(t) \vert + \rangle_N \vert 0 \rangle %
+ \beta_-(t) \vert - \rangle_N \vert 0 \rangle %
+ \cdots %
+ \sum_\mathbf{k} \gamma_\mathbf{k}(t) \vert b_1\cdots b_N \rangle \vert 1_\mathbf{k} \rangle %
\end{align} %
then evolves according to the Hamiltonian
\begin{align} %
\nonumber %
V(t) = \sum_\mathbf{k} \sum_{j=1}^N \hbar g_\mathbf{k} \hat{a}_\mathbf{k} \hat{\sigma}_j^+ e^{i(\omega - \nu_\mathbf{k}) t + i \mathbf{k} \cdot \mathbf{r}_j} + \text{adj.} %
\end{align} %
as follows %
\begin{align} %
\label{Eq:Decay_Small_betaPlus} %
\dot{\beta}_+ = & \, - \frac{i}{\sqrt{N}} \sum_\mathbf{k} \sum_{j=1}^N g_\mathbf{k} e^{i\Delta_\mathbf{k} t + i \mathbf{k} \cdot \mathbf{r}_j} \gamma_\mathbf{k} \\ %
\label{Eq:Decay_Small_betaMinus} %
\dot{\beta}_- = & \, - \frac{i}{\sqrt{N}} \sum_{\mathbf{k}} g_\mathbf{k} e^{i\Delta_\mathbf{k} t} \left( \sum_{j=1}^{N/2} e^{i \mathbf{k} \cdot \mathbf{r}_j}
\quad - \sum_{j'= N/2 + 1}^{N} e^{i \mathbf{k} \cdot \mathbf{r}_{j'}} \right) \gamma_\mathbf{k} \\ %
\label{Eq:Decay_Small_gamma} %
\dot{\gamma}_\mathbf{k} = & \, - i g_\mathbf{k} e^{-i\Delta_\mathbf{k}t} \left\{ %
\frac{1}{\sqrt{N}} \sum_{i=1}^N e^{-i\mathbf{k}\cdot\mathbf{r}_i} \beta_+ %
+ \frac{i}{\sqrt{N}} \left( \sum_{j=1}^{N/2} e^{-i \mathbf{k} \cdot \mathbf{r}_j}
- \sum_{j' = N/2 + 1}^N e^{-i \mathbf{k} \cdot \mathbf{r}_{j'}} \right) \beta_- \right\} %
\end{align} %
where $\Delta_\mathbf{k} = c\vert \mathbf{k} \vert - \omega$. %

Going to the small sample limit, $\mathbf{k}\cdot \mathbf{r}_j \rightarrow 0$, %
we see from Eq.~\eqref{Eq:Decay_Small_betaMinus} that $\dot{\beta}_- = 0$ %
and Eq.~\eqref{Eq:Decay_Small_betaPlus} together with Eq.~\eqref{Eq:Decay_Small_gamma} yields %
\begin{align} %
\dot{\beta}_+ = - \sum_\mathbf{k} \frac{g_\mathbf{k}^2}{N} \int dt' \, e^{i\Delta_\mathbf{k} (t-t')} %
\sum_{i,j=1}^N e^{i\mathbf{k}\cdot(\mathbf{r}_j - \mathbf{r}_i)} \beta_+ (t'). %
\end{align} %
In the small sample limit the $\sum_{i,j}$ term is $N^2$ and the usual limit in which the $\sum_\mathbf{k} e^{i\Delta_\mathbf{k}(t-t')} \Rightarrow \delta(t-t')$ %
we have the Weisskopf-Wigner result $\dot{\beta}_+ = - N \gamma \beta_+$. %

\setcounter{equation}{0} %
\renewcommand{\theequation}{\thesection.\arabic{equation}b} %

Proceeding to the large sample limit we write the state vector
\begin{align} %
\label{Eq:Decay_Large_State} %
\vert \psi(t) \rangle = \beta_+(t) \vert + \rangle_{\mathbf{k}_0} \vert 0 \rangle %
+ \beta_{-}(t) \vert - \rangle_{\mathbf{k}_0} \vert 0 \rangle %
+ \dots %
+ \sum_\mathbf{k} \gamma_\mathbf{k}(t) \vert b_1 \dots b_N \rangle \vert 1_\mathbf{k} \rangle. %
\end{align} %
The equations of motion for the probability amplitudes are now %
\begin{align} %
\label{Eq:Decay_Large_betaPlus} %
\dot{\beta}_+ = - \frac{i}{\sqrt{N}} \sum_\mathbf{k} \sum_{j=1}^N g_\mathbf{k} e^{i\Delta_\mathbf{k} t + i ( \mathbf{k} - \mathbf{k}_0 ) \cdot \mathbf{r}_j} \gamma_\mathbf{k} %
\end{align}
\begin{align} %
\label{Eq:Decay_Large_betaMinus} %
\dot{\beta}_- = - \frac{i}{\sqrt{N}} \sum_\mathbf{k} g_\mathbf{k} e^{i\Delta_\mathbf{k} t} \left[ \sum_{j=1}^{N/2} e^{i ( \mathbf{k} - \mathbf{k}_0 ) \cdot \mathbf{r}_j} %
- \sum_{j'=N/2+1}^N e^{i(\mathbf{k} - \mathbf{k}_0) \cdot \mathbf{r}_{j'}} \right] \gamma_\mathbf{k} %
\end{align} %
\begin{align} %
\nonumber %
\dot{\gamma}_\mathbf{k} = & - i g_\mathbf{k} e^{-i\Delta_\mathbf{k} t} \Bigg\{ %
&& \hspace{-1em}\hspace{-1em} \frac{1}{\sqrt{N}} \bigg[ \sum_{i=1}^N e^{-i(\mathbf{k} - \mathbf{k}_0) \cdot \mathbf{r}_i} \bigg] \beta_+ \\ %
\label{Eq:Decay_Large_gamma} %
&&&  %
\hspace{-1em}\hspace{-1em}
+ \frac{1}{\sqrt{N}} \bigg[ \sum_{j=1}^{N/2} e^{-i(\mathbf{k}-{\mathbf{k}_0})\cdot \mathbf{r}_j} - \sum_{j'=N/2+1}^N e^{-i(\mathbf{k}-{\mathbf{k}_0}) \cdot \mathbf{r}_{j'}} \bigg] \beta_- %
+ \cdots \Bigg\}, %
\end{align}
where the equations are now numbered \eqref{Eq:Decay_Large_State} $\cdots$ \eqref{Eq:DecaySub_Large_betaPlus} to emphasize the connection with the small sample case. %
Integrating Eq.~\eqref{Eq:Decay_Large_gamma} and inserting into Eq.~\eqref{Eq:Decay_Large_betaPlus} yields
\begin{align} %
\nonumber %
\dot{\beta}_+ = & - \int dt' \sum_\mathbf{k} g^2_\mathbf{k} e^{ic( \vert \mathbf{k} \vert - \vert \mathbf{k}_0 \vert ) (t-t')} %
\sum_{j=1}^N \frac{e^{i(\mathbf{k}-{\mathbf{k}_0}) \cdot \mathbf{r}_j}}{\sqrt{N}} \\ %
\label{Eq:DecaySub_Large_betaPlus} %
& \times \bigg\{ \sum_{i=1}^N e^{-i(\mathbf{k} - \mathbf{k}_0) \cdot \mathbf{r}_i} \frac{\beta_+(t')}{\sqrt{N}} %
+ \Big[ \sum_{i=1}^{N/2} e^{-i(\mathbf{k}-{\mathbf{k}_0})\cdot{\mathbf{r}_i}} - \sum_{i'=N/2+1}^N e^{-i(\mathbf{k}-{\mathbf{k}_0}) \cdot \mathbf{r}_{i'}} \Big] %
\frac{\beta_{-}(t')}{\sqrt{N}} + \cdots \bigg\}. 
\end{align}
\renewcommand{\theequation}{\thesection.\arabic{equation}} %
Introducing the notation $\mathbf{K} = \mathbf{k} - \mathbf{k}_0$, noting that
\begin{align} %
\label{A6} %
\sum_{j=1}^N e^{i\mathbf{K}\cdot\mathbf{r}_j}\sum_{j'=1}^N e^{-i\mathbf{K}\cdot \mathbf{r}_{j'}} %
& = \sum_{j=1}^N \left[1+e^{i\mathbf{K}\cdot\mathbf{r}_j} %
\sum_{j' = 1 (\neq j)}^N e^{-i\mathbf{K}\cdot \mathbf{r}_{j'}} \right] %
= N + N (N-1) \frac{(2\pi)^3}{V} \delta(\mathbf{K}) %
\end{align}
and that because $\sum_j e^{i\mathbf{K}\cdot\mathbf{r}_j}\Rightarrow\delta(\mathbf{K})$ %
the coefficients of $\beta_{-}$ etc.~ %
in Eq.~\eqref{Eq:DecaySub_Large_betaPlus} vanish; %
we are left with %
\begin{align}{\label{A7}}
\begin{split}
\dot{\beta}_+=-\int dt'\sum_\mathbf{k} g^2_\mathbf{k} e^{ic(|\mathbf{k}|-|{\mathbf{k}_0}|)(t-t')}\big[1+\frac{(2\pi)^3}{V}(N-1)\delta(\mathbf{k}-\mathbf{{\mathbf{k}_0}})\big]\beta_+(t')
\end{split}
\end{align}
Likewise the equation of motion for $\beta_{-}$ is given by
\begin{align} %
\nonumber %
\dot{\beta}_- = & - \int dt' \sum_\mathbf{k} g^2_\mathbf{k} e^{ic(|\mathbf{k}|-\vert\mathbf{k}_0\vert)(t-t')} %
\bigg[ \sum_{j=1}^{N/2} e^{i(\mathbf{k} - \mathbf{k}_0) \cdot \mathbf{r}_j} - \sum_{j'=N/2+1}^N e^{i(\mathbf{k} - \mathbf{k}_0) \cdot \mathbf{r}_{j'}} \bigg] %
\frac{1}{\sqrt{N}} \\ %
\nonumber %
& \qquad \quad \times \Bigg\{ \sum_{i=1}^N e^{-i(\mathbf{k}-{\mathbf{k}_0}) \cdot {\mathbf{r}_i}} \frac{\beta_+(t')}{\sqrt{N}} \\ %
\label{Eq:Eqn_betaMinusSub} %
& \qquad \qquad \quad + \bigg[ \sum_{i=1}^{N/2} e^{-i(\mathbf{k}-{\mathbf{k}_0}) \cdot \mathbf{r}_i} - \sum_{i'=N/2+1}^N e^{-i(\mathbf{k}-{\mathbf{k}_0}) \cdot \mathbf{r}_{i'}} \bigg] %
\frac{\beta_{-}(t')}{\sqrt{N}} + \cdots \Bigg\} %
\end{align}
and noting that
\begin{align} %
\nonumber %
& \left[\sum_{j=1}^{N/2} e^{i\mathbf{K}\cdot\mathbf{r}_j}-\sum_{j'=N/2+1}^N e^{i\mathbf{K}\cdot\mathbf{r}_{j'}}\right] %
\left[\sum_{i=1}^{N/2} e^{-i\mathbf{K}\cdot\mathbf{r}_i} - \sum_{i'=N/2+1}^N e^{-i\mathbf{K}\cdot \mathbf{r}_{i'}}\right] \\ %
\nonumber %
= & 2 \sum_{j=1}^{N/2} \left[1+e^{i\mathbf{K}\cdot\mathbf{r}_j} \sum_{i = 1 (\neq j)}^{N/2} e^{-i\mathbf{K}\cdot{\mathbf{r}_i}}\right]
- \left[ \sum_{j=1}^{N/2} e^{i\mathbf{K}\cdot\mathbf{r}_j} \sum_{i'=N/2+1}^N e^{-i\mathbf{K}\cdot \mathbf{r}_{i'}} + \text{adj.} \right] \\ %
\label{A9} %
= & N \left[ \left\{ 1 - \frac{(2\pi)^3}{V}\delta(\mathbf{K}) \right\} + \frac{(2\pi)^3}{V} \left( \frac{N}{2}-\frac{N}{2} \right) \delta(\mathbf{K}) \right], %
\end{align}
while the coefficients multiplying $\beta_3, \cdots, \beta_N$ vanish, Eq.~\eqref{Eq:Eqn_betaMinusSub} reduces to
\begin{align} %
\label{Eq:Eqn_betaMinusSubFinal} %
\begin{split}
\dot{\beta}_{-}=-\int dt' \sum_\mathbf{k} g^2_\mathbf{k} e^{ic(|\mathbf{k}|-|{\mathbf{k}_0}|)(t-t')} %
\left[ \left\{ 1 - \frac{(2\pi)^3}{V}\delta(\mathbf{k} - \mathbf{k}_0) \right\} %
+ \frac{(2\pi)^3}{V} \left( \frac{N}{2} - \frac{N}{2} \right) \delta(\mathbf{k}-\mathbf{{\mathbf{k}_0}})\right]{\beta}_{-}(t')
\end{split}
\end{align}
In order to simplify Eq.~(\ref{A7}) and \eqref{Eq:Eqn_betaMinusSubFinal} %
one can write the delta function in terms of the magnitude of $\mathbf{k}$ and solid angle unit vectors $\hat{\Omega}_\mathbf{k}$, etc.~as
\begin{align}{\label{A11}}
\delta(\mathbf{k}-\mathbf{{\mathbf{k}_0}}) = \delta( k - k_0) \delta(\hat{\Omega}_\mathbf{k}-\hat{\Omega}_{{\mathbf{k}_0}})/k^2
= \left[\frac{1}{2\pi}\int\limits_{-R}^{R}e^{i( k - k_0 ) r} dr \right] \delta(\hat{\Omega}_\mathbf{k}-\hat{\Omega}_{{\mathbf{k}_0}}) / k^2,
\end{align}
where $R$ is the radius of the atomic cloud, which is taken to be large compared to $\lambda$ but small compared to the size of the radiation packet size $ct$ where $t\sim1/\gamma$; %
then Eq.~(\ref{A7}) becomes
\begin{align} %
\label{A12} %
\dot{\beta}_+(t) = & \, -\int dt' \sum_\mathbf{k} g_\mathbf{k}^2 e^{-ic(k- k_0)(t-t')} %
\, \left[1+ (N-1) \frac{(2\pi)^2}{Vk^2}\int\limits_{-R}^{R}e^{i(k-k_0)r}dr \, \delta(\hat{\Omega}_\mathbf{k}-\hat{\Omega}_{{\mathbf{k}_0}}) \right]\beta_+(t') %
\end{align}
Changing the sum over \textbf{k} to an integral via the density of states  $V/ \lambda^3$ so that\\ $ \sum_{\mathbf{k}}=\int \frac{V}{(2\pi)^3}k^2dk \sin \theta_k d\theta_k d\varphi_k $, %
Eq.~(\ref{A12}) reads
\begin{align} %
\label{A13} %
\dot{\beta}_+ = & \, -\int dt'\frac{V}{(2\pi)^3}g^2_{\mathbf{k}_0}\int k^2 dk e^{-ic( k - k_0 )(t-t')} %
\, \left[4\pi+ (N-1) \frac{(2\pi)^2}{Vk^2}\int\limits_{-R}^{R}e^{i(k-k_0)r}dr \right]\beta_+(t') %
\end{align}
carrying out the integral over $k$ using $\int e^{-ic(k - k_0)(t-t')}dk = (2\pi/c) \delta(t-t')$ %
we obtain in the first term of Eq.~(\ref{A13}) and $(2\pi/c) \delta(t-t'-r/c) $ in the second. %
For the present purposes we take $ r\ll ct$ and find
\begin{align}{\label{A14}}
\dot{\beta}_+=-\frac{V}{(2\pi)^3}g^2_{\mathbf{k}_0} k_0^2 %
\left[\frac{4\pi^2}{c}+\frac{(N-1)(2\pi)^3}{Vck_0^2} R \right]\beta_+
\end{align}
Tracking the factors of $2\pi$ we write: $V = 4 \pi R^3 / 3$, $k_0^2 = (2\pi)^2 / \lambda^2$
\begin{equation*}
\dot{\beta}_+=-\frac{\pi}{2}\left(\frac{V{k^2_0}}{\pi^2c}\right)g^2_{{\mathbf{k}_0}}\left[1+\frac{2\pi(N-1)\lambda^2}{4\pi \frac{R^3}{3} (2\pi)^2} R \right]\beta_+
\end{equation*}
and defining $ \gamma = \pi \mathscr{D}({\mathbf{k}_0}) g_{{\mathbf{k}_0}}^2 $ %
where $\mathscr{D}(\mathbf{k})={V \mathbf{k}^2}/{\pi^2c}$ so that Eq.~\eqref{A14} finally yields
\begin{align} %
\label{A15} %
\dot{\beta}_+ \cong -\frac{\gamma}{2} %
\left[ 1 + \frac{3}{8\pi} \frac{\lambda^2}{A}(N-1) \right] \beta_+. %
\end{align}
Likewise the equation of motion for $\dot{\beta}_-$, Eq.~\eqref{Eq:Eqn_betaMinusSubFinal}, is
\begin{align} %
\label{A16} %
\dot{\beta}_- \cong -\frac{\gamma}{2} \left[ \left\{ 1 - \frac{3}{8\pi}\frac{\lambda^2}{A} \right\} %
+ \frac{3}{8\pi}\frac{\lambda^2}{A} \left( \frac{N}{2} - \frac{N}{2} \right) \right] \beta_{-}. %
\end{align} %

\section*{Supplement B: State preparation via longitudinal and transverse excitation with single photon quantum and classical fields}
\renewcommand{\thesection}{\Alph{section}} %
\setcounter{section}{2} %
\setcounter{equation}{0} %
\setcounter{figure}{0} %
\numberwithin{equation}{section}
\numberwithin{figure}{section}
\renewcommand{\theequation}{\thesection.\arabic{equation}}
\renewcommand{\thefigure}{\thesection\arabic{figure}} %

\begin{figure}[ht] %
\centering %
\includegraphics[width=1.0\textwidth]{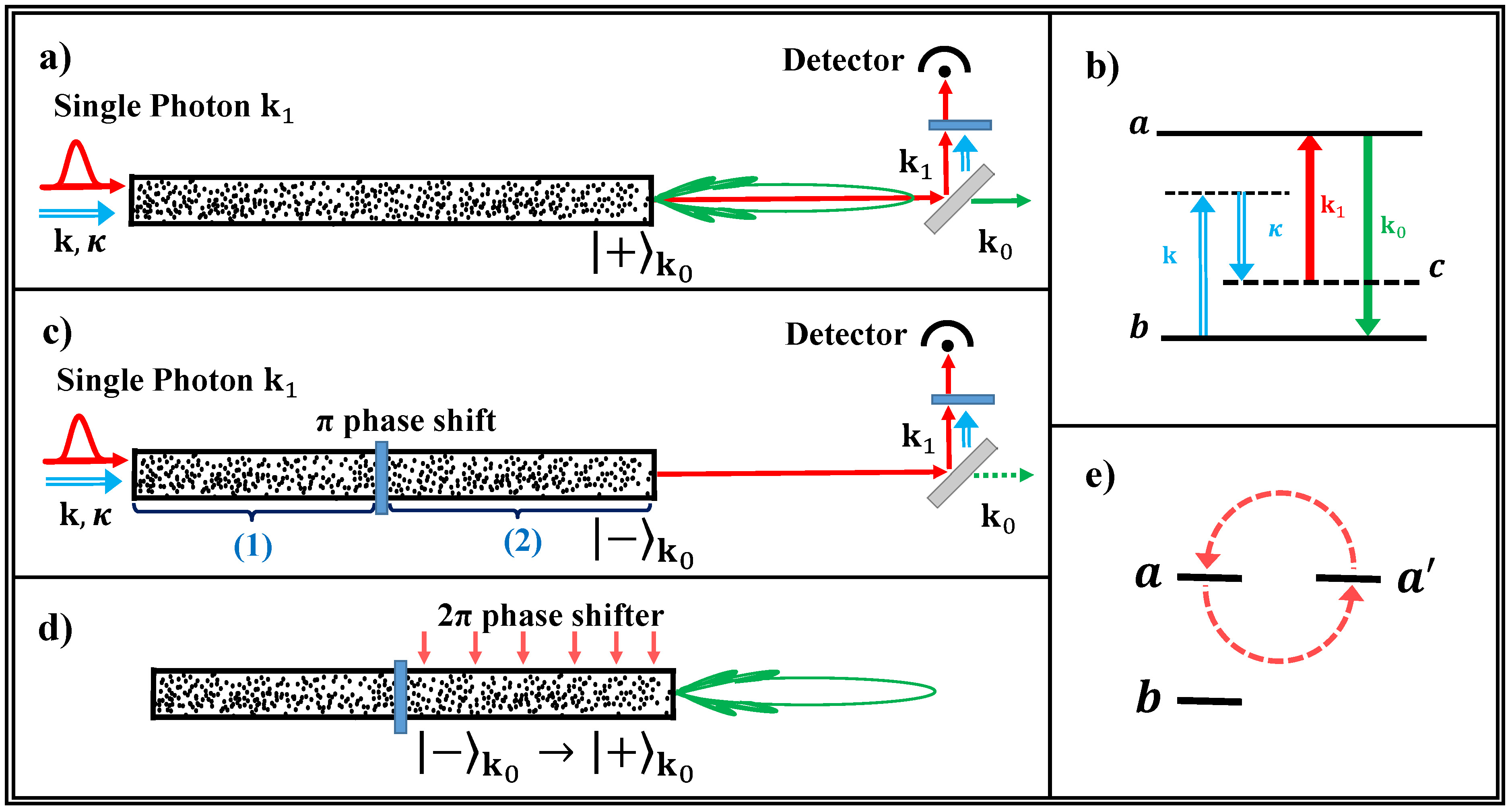}
\caption{a) Single photon of wave vector $\mathbf{k}_1$ is accompanied by a lasers having wave vectors $\mathbf{k}$ and $\boldsymbol{\kappa}$ %
such that $\mathbf{k} - \boldsymbol{\kappa} + \mathbf{k}_1 = \mathbf{k}_0$. %
The $\mathbf{k}_0$ photon is resonant with the transition  $\vert a \rangle$ to $\vert b \rangle$. %
The atoms are weakly driven by the excitation process i.e.~the atomic medium is optically thin during the preparation process, %
and most of the $\mathbf{k}_1$ single photon pulses register a count in the detector; %
the $\mathbf{k}$ and $\boldsymbol{\kappa}$ radiation is isolated from the detector. %
The lack of a count heralds the preparation of the $\vert + \rangle_{\mathbf{k}_0}$ state. %
b) The atoms are driven by a Raman-type process in which two photons $\mathbf{k}$ and $\boldsymbol{\kappa}$ excite the atom to the virtual state $c$ %
and the $\mathbf{k}_1$ photon %
takes the atom to the $\vert a \rangle$ state. %
This panel is to be compared with part c of Fig.~2. %
There we simplified the presentation by taking the state $\vert a \rangle$ to have mixed parity. %
c) The $\vert - \rangle_{\mathbf{k}_0}$ state is prepared by phase shifting the $\mathbf{k}_1$ photon by $\pi$, %
which is the purpose of the $\pi$ phase shifter. %
The $\mathbf{k}_0$ photons are not affected by the phase shifter. %
This results in multiplication of the $\vert a_{j'} \rangle$ atoms on the RHS of phase shifter by $-1$; %
the net result is that a no-count event signals the fact that the $\vert - \rangle_{\mathbf{k}_0}$ state has been prepared. %
d) Cycling the RHS atoms $a \rightarrow a' \rightarrow a$ as per Fig.~2e results in another factor of $-1$ multiplying the region $(2)$ $\vert a \rangle$ atoms %
and takes the single photon subradiant state $\vert - \rangle_{\mathbf{k}_0}$ to $\vert + \rangle_{\mathbf{k}_0}$. %
} %
\label{Fig:SinglePhotonSubradiantState} %
\end{figure}

Two central features of state preparation in single photon superradiance are: %
(1) Optically thin media on preparation but thick on emission. %
(2) Post-selection based on single longitudinal photon detection. %
We here address both points as in Fig.'s ~B1 and B2, %
and also consider classical fields instead of quantized single photon excitation fields. %
Concerning point $1$, transverse time delayed pulses can replace longitudinal excitation, %
where we use $\mathcal{E} e^{-i\nu_0 (t - t_i)}$ where $t_i = z_i / c$ %
or $\mathcal{E} e^{-i\nu_0 t + k_0 z_i}$ since $\nu_0 = c k_0$. %
Thus a single photon divided by beam splitters and sequentially delayed %
will prepare the $\vert + \rangle_{\mathbf{k}_0}$ of Eq.~(3b). %

\begin{figure}[ht] %
\centering
\includegraphics[width=0.6\textwidth]{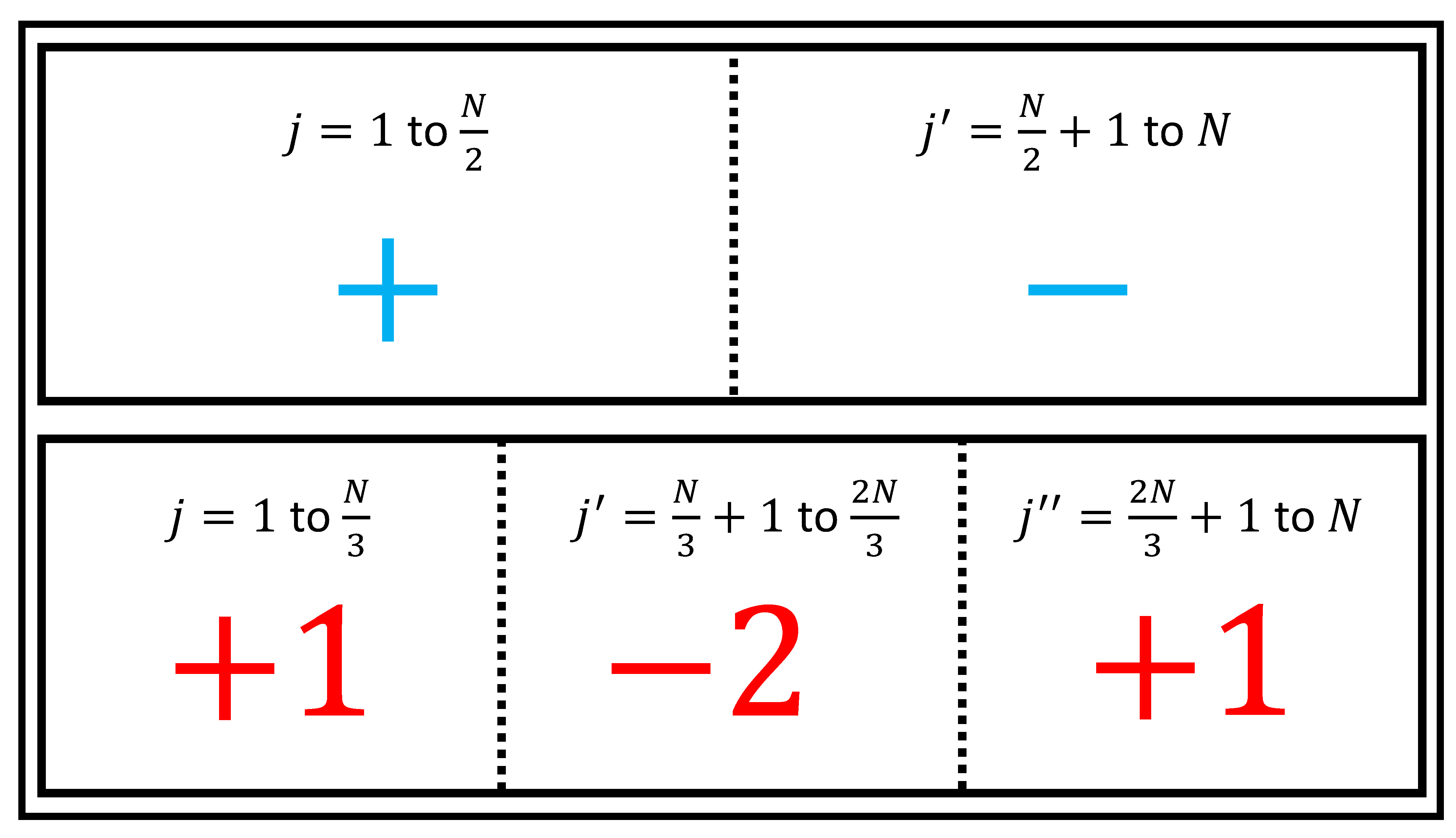} %
\caption{The subradiant states $\vert - \rangle$ and $\widetilde{\vert - \rangle}$ divide atoms into spatially separated subensembles. %
The upper panel is associated with the many particle state of Fig.~2 and Eq.~(3b) of the text. %
The lower panel clarifies the notation associated with Eq.~(9). %
The red numerals correspond to the coefficients in Eq.~(B.2a). %
} %
\label{Fig:PreparationThreeState} %
\end{figure}

The subradiant state is best understood by explicitly writing it out %
and symmetrizing in the second to obtain %
\begin{subequations} %
\begin{align} %
\nonumber %
\vert - \rangle_{\mathbf{k}_0} = & \, \frac{1}{\sqrt{2N_2}} \frac{1}{N_2} \sum_{j=1}^{N_2} \sum_{j'=N_2+1}^{2N_2} \Big[ %
\vert b_1 \cdots a_j \cdots b_{N_2} \rangle \vert b_{N_2+1} \cdots b_{j'} \cdots b_{N} \rangle e^{i\mathbf{k}_0 \cdot \mathbf{r}_j} \\ %
& \qquad \qquad %
- \vert b_1 \cdots b_j \cdots b_{N_2} \rangle \vert b_{N_2+1} \cdots a_{j'} \cdots b_{N} \rangle e^{i\mathbf{k}_0 \cdot \mathbf{r}_{j'}} %
\Big] \\ %
= & \, \frac{1}{\sqrt{N_2}} {\sum_{j,j'}}^{2} \frac{1}{\sqrt{2}} \Big[ \vert a_j b_{j'} \rangle - \vert b_j a_{j'} \rangle \Big] \vert \{ b \}_{jj'} \rangle, %
\end{align} %
\end{subequations} %
where $\vert \{ b \}_{jj'} \rangle$ is the state will all atoms in the lower level but $j$ and $j'$ are omitted. %

Other single photon subradiant states can be similarly prepared as in Fig.~B2. %
There we sketch three subensembles (atomic bins) and the atoms are labeled by $j$, $j'$, $j''$ for subensembles $1$, $2$, $3$; %
each containing $N_3$ atoms. %
This kind of many particle subradiant state is given by %
\begin{subequations} %
\begin{align} %
\label{Eq:GeneralizedSubradiant} %
{\vert 3 \rangle}_{\mathbf{k}_0} = \frac{1}{\sqrt{6}} %
\left[ \sum_{j=1}^{N/3} \frac{e^{i{\mathbf{k}_0} \cdot\mathbf{r}_j}}{\sqrt{N_3}} \vert j \rangle %
-2 \sum_{j'=N/3+1}^{2N/3} \frac{e^{i{\mathbf{k}_0}\cdot\mathbf{r}_{j'}}}{\sqrt{N_3}} \vert j' \rangle %
+ \sum_{j''=2N/3+1}^N \frac{e^{i{\mathbf{k}_0}\cdot\mathbf{r}_{j''}}}{\sqrt{N_3}} \vert j'' \rangle \right]\ ,\
\end{align} %
where in the setup of Fig.~B2 the number of atoms in each zone is $N_3 = N/3$, %
the state $\vert 3 \rangle_{\mathbf{k}_0}$ is called $\widetilde{\vert - \rangle}$ in Eq.~(12) to emphasize the connection with the $\vert - \rangle$ state of Eq.~(1b). %

The state $\vert 3 \rangle_{\mathbf{k}_0}$ of Eq.~\eqref{Eq:GeneralizedSubradiant} decays according to Weisskopf-Wigner (WW) theory at the rate %
\begin{align} %
\label{Eq:DecayRate_Subradiant} %
\Gamma_{3,N} \cong %
\frac{\gamma}{2} \left[ \left( 1 - \frac{3}{8\pi} \frac{\lambda^2}{A} \right) %
+ \frac{3}{8\pi} \frac{\lambda^2}{A} \left( \frac{N}{3} - \frac{N}{3} \right) \right], %
\end{align} %
\end{subequations} %
as can be shown using the approach of Supplement A. %
Note that in this application of WW theory $\vert 3 \rangle_{\mathbf{k}_0}$ %
this state is stable for times small compared to ${1}/{\gamma }$. %
Physically the photon is stored in the atomic ensembles represented by $\vert - \rangle_{\mathbf{k}_0}$ and $\vert 3 \rangle_{\mathbf{k}_0}$ %
by arranging the atomic oscillators to be properly phased. %
That is, in the case of $|-\rangle_{{\mathbf{k}_0}}$ the atomic antennas in the $j$ and $j'$ subensembles are $\pi$ out of phase corresponding to the minus sign in Eq.~(3b). %
In the subradiant state Eq.~\eqref{Eq:GeneralizedSubradiant} the emissions from the ${j}$ and ${j'}$ subensembles are $\pi$ out of phase %
with the emission from the ${j''}$ subensemble atoms which are excited with twice the amplitude. %

Point 2 of the introductory remarks of this Supplement calls for more discussion. %
For the present purposes it suffices to use a small sample such that we are in the Dicke limit $\lambda_0 > R$. %
In transverse excitation tight focus can be achieved %
`if we envision Raman type' excitation as in e.g., Fig.~B1b. %
since the Raman fields can have much shorter wavelength than $\lambda_0 = 2 \pi / k_0$. %
Then we may in effect address individual atoms (e.g.~in an ion trap or thin glass rod), as in Fig.'s~3, 4 %
of the text. %

\end{appendices}